\documentstyle[twoside,fleqn,espcrc2,epsfig]{article}

\title{Chiral symmetry in the 2-flavour lattice 
Schwinger model\thanks{Supported by Fonds zur F\"orderung der 
Wissenschaft\-li\-ch\-en Forschung in \"Osterreich, 
Project P11502-PHY.}}      

\author{I. Hip\address{Institut f\"ur Theoretische Physik,
Universit\"at Graz, A-8010 Graz, Austria},
C.~B. Lang$^{\rm a}$ and \phantom{\footnotemark}
R. Teppner$^{\rm a}$\thanks{presented by R. Teppner}}

\begin{document}

\begin{abstract}
We study the 2-flavour lattice Schwinger model: QED in $D=2$ with
two fermion species of identical mass.  In the simulation we are using
Wilson fermions where chiral symmetry is explicitly broken.  Since
there is no known simple order parameter it is non-trivial to identify
the critical line of the chiral phase transition.  We therefore  need
to find observables which allow an identification of a possible
restoration of chiral symmetry.  We utilize the PCAC-relations in order
to identify the critical coupling, where chiral symmetry is restored.
\end{abstract}

\maketitle

\section{INTRODUCTION}

The 2-flavour Schwinger model is a 2D theory of two fermion species and
photons. In the massless continuum model the $SU(2)_A$ flavour symmetry
is broken by a mechanism mimicking spontaneous symmetry breaking.
Due to the Mermin-Wagner-Coleman theorem \cite{MeWaCo66} in
this situation there is no spontaneous symmetry breaking and thus there
are no Goldstone bosons, nevertheless massless pions are expected from
continuum studies (for a recent study cf. \cite{GaSe94}).

The naive lattice formulation for fermions shows the notorious doubling
problem. In Wilson's action a term is added giving the doublers a mass
${\cal O}(1/a)$ decoupling them in the continuum limit.  Independent of
the possible spontaneous breaking of chiral symmetry (which does not
occur in 2D but is expected in e.g. QCD$_4$) we expect that the chiral
symmetry of the action itself is restored at a critical fermion
coupling $\kappa_c(\beta)$. However, since the Wilson term breaks
chiral symmetry explicitly we have no known simple order parameter,
which allows one to identify that position.  Following earlier
proposals \cite{BoMaMa85} it has recently been suggested to utilize
PCAC-relations with Schr\"odinger functional methods for that purpose
\cite{JaLiLu96}; this turned out to be quite powerful in the context of
QCD$_4$ \cite{Lu97}.  Here we report on results for QED$_2$ with 2
species of fermions, where we also employed PCAC-relations in order to
determine the critical fermion coupling line.

The lattice action for the massive 2-flavour Schwinger model with Wilson
fermions is
\begin{equation}
S[U,\psi,\bar\psi]=S_{g}[U]+S_{f}[U,\psi,\bar\psi]\;,
\end{equation}
where $S_g$ is the compact plaquette gauge action (coupling $\beta$),
$\psi=(u,d)$ is a flavour-doublet and
\begin{equation}
\begin{array}{l}
\displaystyle
S_{f}[U,\psi,\bar\psi]=\sum_{n}\bar\psi_{n}\psi_{n} \\
\displaystyle
\quad-\kappa\sum_{n,\mu}\left[\bar\psi_{ n}\,
(r-\sigma^{ \mu})\,U_{n,\mu}\,\psi_{n+\mu}+h.c.\right]\;.
\end{array}
\end{equation}
The $\bar\psi\psi$ term and the terms proportional to $r$ 
(we choose $r=1$) break chiral symmetry explicitly. 
Strictly speaking the action has only the $SU(2)_V\times U(1)_V$ global
symmetries. The main idea using the PCAC-concept is to find the points in the
coupling space, where the explicit breaking of the chiral symmetries
$SU(2)_A\times U(1)_A$  vanishes.  The fermions were simulated with the
HMC-method; more details can be found in \cite{GaHiLa97}.

\section{WARD IDENTITIES}

Global symmetries of an action on the classical level lead to conserved
Noether-currents. In a QFT the symmetries may be spontaneously broken
or broken due to anomalies resulting from the functional integration.
So the relations expressing current conservation may change depending
on the mechanism of symmetry violation.  To derive these relations one
considers global infinitesimal symmetry transformations of a path
integral
\begin{equation}
     \langle 0|{\cal O}[U, \bar\psi, \psi ]|0 \rangle=\frac{1}{Z}\int dU d\bar\psi d\psi
     {\cal O}[U, \bar\psi, \psi ] e^{-S},
\end{equation}
where $ {\cal O}$ denotes an arbitrary operator. If the integration measure is invariant
this leads to relations like
\begin{equation}
\langle 0|\delta {\cal O}|0\rangle-\langle 0|{\cal O}\delta
S|0\rangle=0\;,
\end{equation}
called Ward identities. Thus, the quantum analog to the classical
conservation law has similar (or equal) form but is an operator
identity.

\begin{figure}[t]
\begin{center}
\epsfig{file=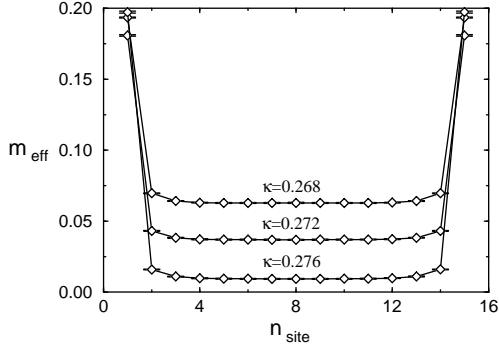,angle=270,width=6.5cm}
\end{center}
\vspace{-24pt}
\caption{Mass plateaus at $\beta=2.0$ for the $16\times 16$ lattice
(10000 configurations).}
\vspace{-9pt}
\label{fig1}
\end{figure}

For the massive continuum  Schwinger model with two flavours the 
corresponding identities
for the global $SU(2)_V\times U(1)_V$ symmetries which are exactly
realized in the QFT are
\begin{equation}
\bullet \;\;  SU(2)_{V} \quad 
\partial_{\mu}\,j^{\mu  a}=0\;, \;\;
 j^{\mu  a}=\bar\psi  \,\frac{\tau^{a}\otimes\sigma^{\mu}}{2}\,\psi\;,
\end{equation}
\vspace{-8pt}
\begin{equation}
\bullet \;\; U(1)_{V}\phantom{S} \quad 
\partial_{\mu}\,j^{\mu}\;=0\;, \;\;
j^{\mu}=\bar\psi \, \frac{1_{2\times 2}
\otimes\sigma^{\mu}}{2}\,\psi\;.
\end{equation}
The $SU(2)_A\times U(1)_A$ symmetries are broken for massive fermions.
The $SU(2)_A$ Ward identity (PCAC-relation) reads
\begin{eqnarray}\label{WISU2A}
\bullet \;\;  SU(2)_{A} \quad
\partial_{\mu} \,j^{\mu a 3}=
2\,m_{0}\,\pi^a\;,
\pi^a=\bar\psi\,\frac{\tau^{a}\otimes
\sigma^{3}}{2}\,\psi\;,\quad
\nonumber
\end{eqnarray}
\vspace{-9pt}
\begin{equation}
\phantom{\bullet \;\;  SU(2)_{A} \quad}
j^{\mu a 3}=
\bar\psi\,\frac{\tau^{a}\otimes\sigma^{\mu}\sigma^{3} }{2}\,\psi\;.
\end{equation}
The $U(1)_A$ symmetry is also broken by the anomaly
\begin{eqnarray} \nonumber
\bullet  \;\;  U(1)_{A}\;\;
 \partial_{\mu}\, j^{\mu 3}=2\,m_{0} \,
 \bar\psi\,\frac{1_{2\times 2}\otimes \sigma^{3}}{2}\,\psi\,+\,
\mbox{anomaly}\,,
\nonumber     
\end{eqnarray}     
\vspace{-12pt}
\begin{equation}   
\phantom{\bullet \;\;  U(1)_{A} \;\;}
j^{\mu 3}=\bar\psi \, \frac{1_{2\times
2}\otimes\sigma^{\mu}\sigma^{3}}{2}\,\psi\,.
\end{equation}

\begin{figure}[t]
\begin{center}
\epsfig{file=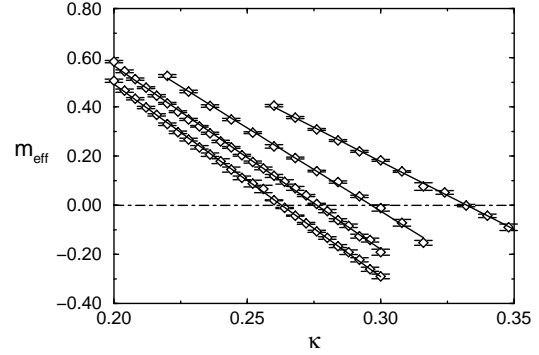,angle=270,width=7cm}
\vspace{-24pt}
\end{center}
\caption{Values (and linear fits) obtained for $\hat m_{\textrm{\footnotesize eff}}$ for 
different $\kappa$ at 
a given $\beta$ (from left to right: 0.1, 1, 2, 4) 
on an $8\times 8$ lattice (1000 configurations each).}
\vspace{-9pt}
\label{fig2}
\end{figure}

\section{RESULTS}

The underlying concept is to utilize a lattice version of
(\ref{WISU2A}) in order to identify the value of $\kappa_{c}(\beta)$
where $SU(2)_A$ holds.  That point defines a theory with massless
lattice fermions.  The naively discretized continuum action
\begin{eqnarray}
S=S_g[U]+
\hat m_{\textrm{\footnotesize eff}}\sum_{n}\bar\psi_{ n}\psi_{n} \nonumber 
\end{eqnarray}
\vspace{-9pt}
\begin{eqnarray}
-\frac{1}{2}\sum_{n \mu}\{\bar\psi_{n+\mu}\sigma^{\mu}
U^{+}_{n,\mu}\psi_{ n}+
\bar\psi_{n}(-\sigma^{ \mu})U_{n,\mu}\psi_{n+\mu}\}
\end{eqnarray}
leads to an $SU(2)_A$ lattice operator identity analogous to 
(\ref{WISU2A}). In order to satisfy that relation non-trivially we
choose to take expectation values with regard to a pion source
and obtain
\begin{eqnarray}
\forall n\not=y :\;\; 2\,\hat 
m_{\textrm{\footnotesize eff}}\,\langle 0|\,\pi^{a +}_{y}\,\pi^{a}_{n}\,|0\rangle=
\nonumber
\end{eqnarray}
\vspace{-8pt}
\begin{eqnarray}\label{meffdef}
\qquad
\sum_{\mu}\left[\langle 0|\,\pi^{a +}_{y}\,\hat J^{3 a \mu}_{n}\,|0\rangle -
\langle 0|\,\pi^{a +}_{y}\,\hat J^{3 a \mu}_{n-\mu}\,|0\rangle\right]
\end{eqnarray}
with
\begin{eqnarray}
\hat J^{3 a \mu}_{n}=\frac{1}{2}
\left[\bar\psi_{ n+\mu}\,U^{+}_{n,\mu}
\frac{\tau^{a}\otimes\sigma^{\mu}\sigma^{3}}{2}\,\psi_{n}+\right.
\nonumber
\end{eqnarray}
\vspace{-8pt}
\begin{eqnarray}
\qquad\qquad\left.
\bar\psi_{n}\,\frac{\tau^{a}\otimes\sigma^{\mu}\sigma^{3}}{2}
U_{n,\mu}\,\psi_{n+\mu}\right].
\end{eqnarray}
The operators in the above relation are projected onto states with zero
momentum by summing over time-slices. This relation enables us to
measure the mass parameter $\hat m_{\textrm{\footnotesize eff}}$.  Here we are only interested
in the value $\kappa_c(\beta)$, where $\hat m_{\textrm{\footnotesize eff}}$ vanishes.

\begin{figure}[t]
\begin{center}
\epsfig{file=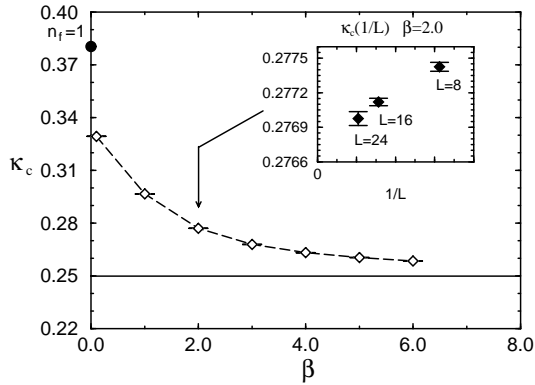,angle=270,width=7cm}
\vspace{-24pt}
\end{center}
\caption{Phase diagram for the 2-flavour model
for the $16\times 16$ lattice (dashed lines to guide the eye); the value 
for the  1-flavour model is from \cite{GaLa95}.}
\label{fig3}
\vspace{-9pt}
\end{figure}

The identity (\ref{meffdef}) should hold independent of the
time-separation between source and sink. In \cite{JaLiLu96} such a
behaviour in the context of the Schr\"odinger functional is used to
determine the value of improvement coefficients. In our case we find
plateaus already as soon as the distance extends beyond the overlap
region of the operators (fig.~1). For the analysis we  excluded the
points up to distance 3 (for $8\times 8$ lattice: 2).

The remaining points represent a mass plateau giving the value of $\hat
m_{\textrm{\footnotesize eff}}$ and its error is obtained from the weighted average.  The
errors of the correlation functions were computed using the jackknife
method.

For given $\beta$ we determined the values of $\hat m_{\textrm{\footnotesize eff}}$ for
different values of $\kappa$.  According to the general idea chiral
symmetry is restored where $\hat m_{\textrm{\footnotesize eff}}$ vanishes.  Interpolating the
values of $\hat m_{\textrm{\footnotesize eff}}$  we get the desired value of $\kappa_c(\beta)$
(fig.~2). In the domain studied a linear fit is appropriate. The
resulting phase diagram is given in fig.~3.

\begin{figure}[t]
\begin{center}
\epsfig{file=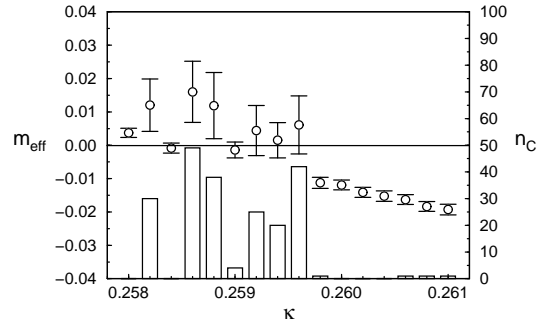,angle=270,width=7cm}
\vspace{-24pt}
\end{center}
\caption{Dependence of $\hat m_{\textrm{\footnotesize eff}}$ (circles) on the
topological charge $\nu$
($\beta=6$, lattice size $8\times 8$, 1000 configurations each; $n_c$ 
denotes the number of configurations with non-zero $\nu$).}
\label{fig4}
\vspace{-8pt}
\end{figure}

Finally, fig.~\ref{fig4} indicates possible problems.  When measuring
$\hat m_{\textrm{\footnotesize eff}}$ we simultaneously kept track of the topological charge
$\nu$ of the configurations. In particular at larger $\beta$ tunneling
between different sectors becomes less frequent \cite{Si97}.  Our
results demonstrate a distinct dependence of the measured $\hat
m_{\textrm{\footnotesize eff}}$ on $\nu$ which may lead to problems in situations where a
simulation is non-ergodic.

\end{document}